\documentclass[letterpaper]{JHEP3}

\usepackage{epsfig}
\usepackage{graphicx}
\usepackage{amssymb,amsmath}
\usepackage{epstopdf}

\def\be{\begin{eqnarray}}
\def\ee{\end{eqnarray}}

\preprint{TAUP-2802/07}

\title{Quark mass and condensate in HQCD}

\author{Oren Bergman\\
Department of Physics\\
Technion, Israel Institute of Technology\\
Haifa 32000, Israel\\
\email{bergman@physics.technion.ac.il}}

\author{Shigenori Seki and Jacob Sonnenschein\\
School of Physics and Astronomy\\
The Raymond and Beverly Sackler Faculty of Exact Sciences\\
Tel Aviv University\\
Ramat Aviv, 69978, Israel\\
\email{sekish,cobi@post.tau.ac.il}}

\abstract{We extend the Sakai-Sugimoto holographic model of QCD (HQCD)
by including the scalar bi-fundamental ``tachyon" field in the 8-brane-anti-8-brane
probe theory.
We show that this field is responsible both for the spontaneous breaking
of the chiral symmetry, and for the generation of (current algebra) quark masses,
from the point of view of the bulk theory.
As a by-product we show how this leads to the Gell-Mann-Oakes-Renner relation
for the pion mass.}

\begin{document}
\maketitle

\section{Introduction}

\begin{figure}
\centerline{\epsfig{file=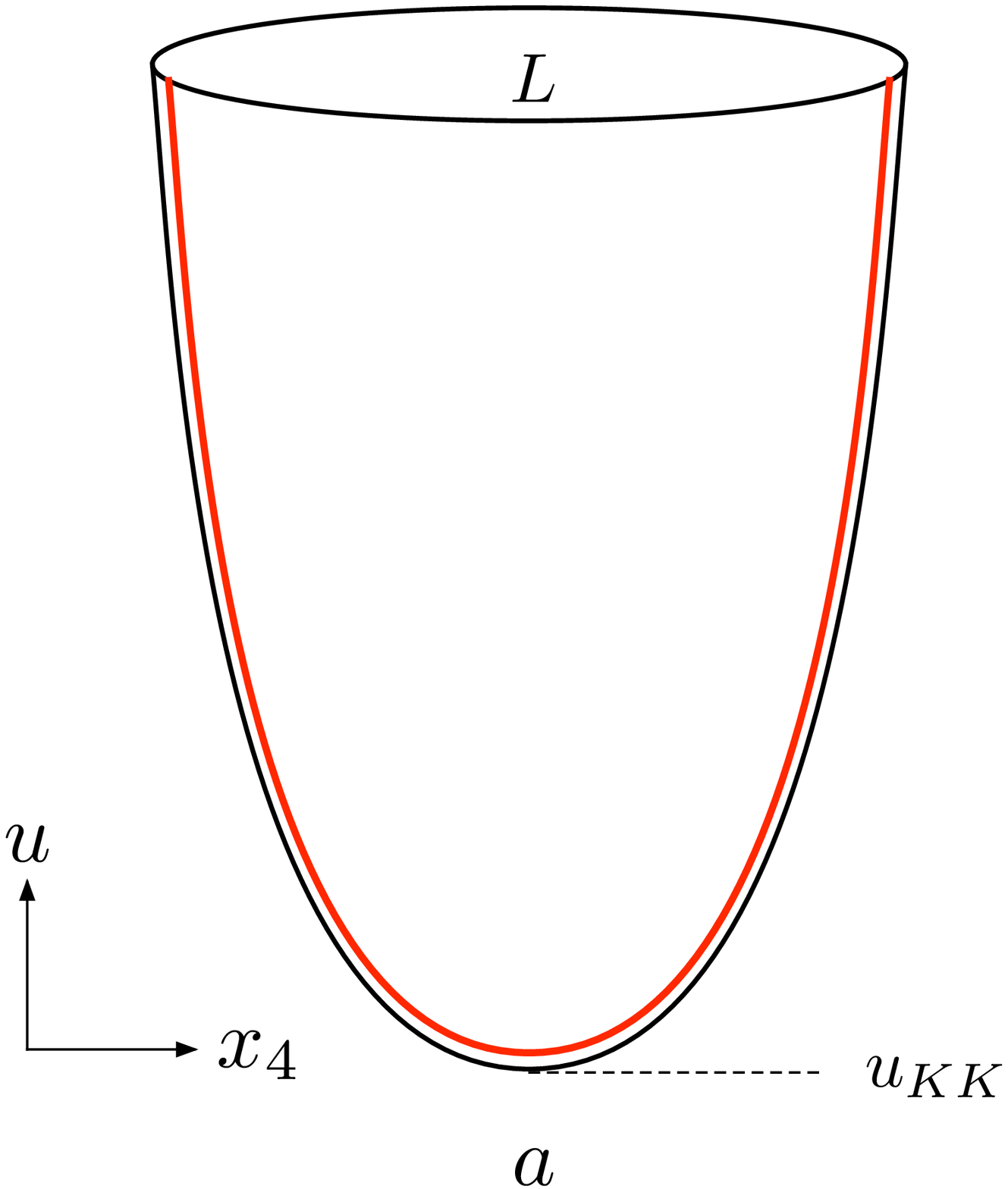,height=4cm}\hspace{1cm}
\epsfig{file=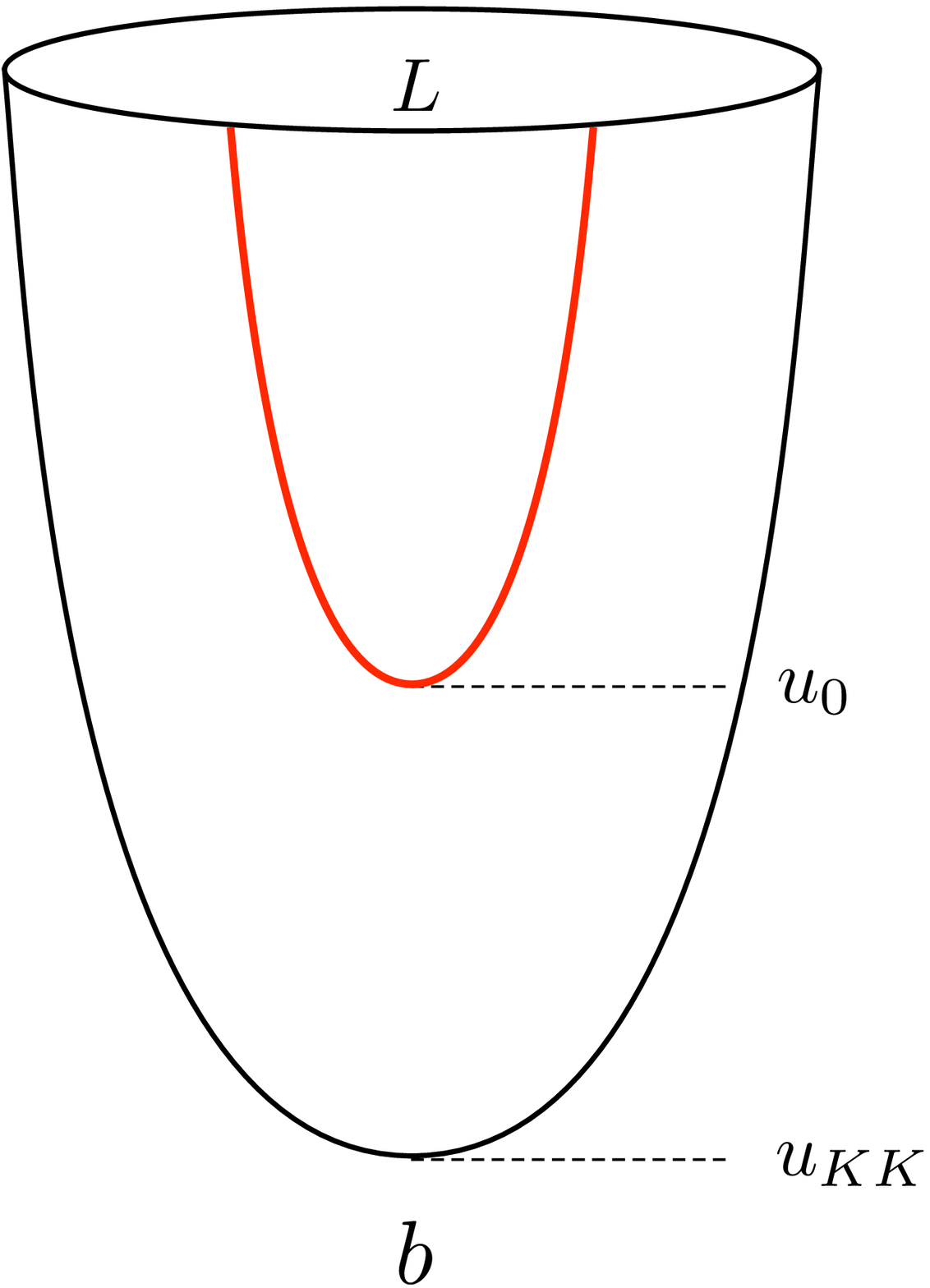,height=4cm}}
\caption{The 8-brane-anti-8-brane configuration in the compact non-extremal 4-brane
background. (a) The anti-podal case. (b) The general case.}
\label{SS}
\end{figure}

The closest model so far to a holographic description of large $N_c$
QCD is the Sakai-Sugimoto model \cite{Sakai:2004cn}. 
The $U(N_c)$ gauge sector is described, at low energy, 
by the near-horizon limit of $N_c$ D4-branes
wrapped on a Scherk-Schwarz circle \cite{Witten:1998zw}. 
This description is valid at energies well below the Kaluza-Klein scale of the circle.
The quark sector  is incorporated by including   $N_f$ D8-branes and
$N_f$ anti-D8-branes transverse to the circle. The
strings that stretch between the original D4-branes and the D8-branes
(anti-D8-branes) describe right-handed (left-handed) chiral fermions, 
which transform in the fundamental representation of both the $U(N_c)$ 
color group and the $U(N_f)_R$ ($U(N_f)_L$) flavor group. 
It is assumed that $N_f\ll N_c$, so that the D8-branes can be treated 
as probes in the D4-brane background, and one can ignore their backreaction.
In QCD this corresponds to the quenched approximation. 
The Sakai-Sugimoto model shares
many features with other holographic models of gauge theory with
matter, however the novel feature of this model is the
geometrical realization of spontaneous  flavor chiral symmetry
breaking. 
The 8-branes and anti-8-branes are separated along the circle
asymptotically in the radial coordinate, but are connected at some 
minimal radial position (figure~\ref{SS}). The former corresponds in QCD to the
UV flavor symmetry being $U(N_f)_R\times U(N_f)_L$,
and the latter corresponds to the IR symmetry being
the diagonal subgroup $U(N_f)_V$.

In spite of its  success, there are  several open questions about
the model, some of which are related to very basic notions of gauge
dynamics. 
The first is the incorporation of a QCD, or ``current algebra'' quark mass.
The quarks are massless in this model since the 8-branes necessarily intersect
the 4-branes.
This is also manifested in the fact that the modes identified with the pions are
massless. It is well known that pions obey the
Gell-Mann-Oakes-Renner relation \cite{GellMann:1968rz}
\be
\label{GOR}
M^2_\pi =  \frac{2m_q \langle\bar q q\rangle}{f_{\pi}^2} + {\cal O}(m_q^2)\,.
\ee 
Therefore in a non-trivial quark condensate massless pions imply massless quarks.
In the original construction of \cite{Sakai:2004cn} the 8-branes and anti-8-branes
were located at antipodal points on the circle, and they connected at the minimal
radial position of the background $u_{KK}$.
This was extended in \cite{Casero:2005se},\cite{Aharony:2006da},\cite{Peeters:2006iu} to a family of configurations parameterized
by the asymptotic separation $L$, or equivalently by the minimal
radial position of the 8-branes $u_0$ (see figure~\ref{SS}(b)).
In these configurations there is a natural mass parameter associated with the
``length'' of a string stretched from the minimal radial position of the 
background $u_{KK}$ to the minimal radial position of the 8-branes $u_0$. 
However, it is easy to check that the pions remain massless,
so this parameter cannot be identified with the current algebra mass.
It is instead related to the ``constitutent quark mass''. 
Indeed in \cite {Peeters:2006iu} for the model of  \cite{Sakai:2004cn}, and in 
\cite{Casero:2005se} for analogous non-critical models,
 it was found that the masses of the vector mesons are linearly related to this
``length of the string'' mass parameter. Moreover, a model of the
decay process of  spinning stringy mesons \cite{Peeters:2005fq}  supports the
interpretation of the ``length'' as the constituent mass.

A related question has to do with the quark condensate itself:
how does one compute it in the holographic description?\footnote{In \cite{Kruczenski:2003uq} both the current algebra mass of the quarks and the condensate can be read 
from the profile of the flavor branes,  and the GOR relation of (\ref{GOR}) is obeyed.
 However that model suffers from the drawback that it does not incorporate chiral flavor symmetry.}
The answers to both questions  are related to each other.
The quark mass term in QCD is
\be
m_q\, \bar{q} q = m_q\,( q^\dagger_R q^{\phantom{\dagger}}_L 
+ q^\dagger_L q^{\phantom{\dagger}}_R)\,,
\ee
where $m_q$ is an $N_f\times N_f$ matrix, $q_R$ is a fundamental
of $U(N_f)_R$, and $q_L$ is a fundamental of $U(N_f)_L$.
Both the mass and the quark bi-linear should therefore be identified
with a bi-fundamental field in the bulk.\footnote{In \cite{Evans:2007jr} 
the quark mass and condensate were
identified with the scalar field corresponding to the 8-brane-anti-8-brane separation.
We believe this is incorrect, since this field
transforms in the adjoint, rather than the bi-fundamental, representation of the 
$U(N_f)$s.}
In this model the required bi-fundamental field comes from 
the D8-$\overline{\mbox{D}8}$ strings.\footnote{The same field appears in 
the holographic description of the resolution of the $U(1)_A$ puzzle
\cite{Bergman:2006xn}.}
The dual operator is therefore non-local in the coordinate transverse to the 8-branes.
According to the usual holographic dictionary the normalizable mode of this field
should correspond to the expectation value of the quark bi-linear, {\em i.e.}
to the quark condensate, and the non-normalizable mode should correspond
to the quark mass.

The bi-fundamental field was not included in the analysis of \cite{Sakai:2004cn},
since, as was argued there, it is very massive.
The 8-brane-anti-8-brane separation was assumed to be much greater than
the string length, which in flat space would make this field massive.
However the proper distance between the 8-branes and anti-8-branes
in the curved background of this model
depends on the radial coordinate $u$, and decreases as $u$ decreases.
The mass of the bi-fundamental field therefore depends on $u$ as well.
For the U-shaped configuration found in \cite{Sakai:2004cn} this field remains massive
for all $u$. While the proper distance decreases as $u$ decreases, the relative angle
between the 8-branes and anti-8-branes increases, so that it never becomes tachyonic.
The situation changes in the case of the non-compact background considered in
\cite{Antonyan:2006vw}. In that case $u_{KK} = 0$, so 
there are two possible configurations: a connected U-shaped
8-brane similar to the one of \cite{Sakai:2004cn}, and a disconnected parallel
8-brane-anti-8-brane configuration. 
In the parallel configuration the proper distance between 
the 8-branes and anti-8-branes goes to zero at $u=0$, and therefore
the bi-fundamental field becomes tachyonic in a finite range of $u$ near the origin.
This represents a (radially) localized tachyonic instability, and one expects the true vacuum to be the U-shaped
configuration\footnote{A similar effects occur in the ``hairpin brane" of 
\cite{Kutasov:2005rr}, and in the meta-stable supersymmetry breaking 
brane configurations of \cite{Giveon:2007fk}.}.
The condensation of this localized tachyon
can be seen as a Higgs-like effect which breaks the chiral symmetry
$U(N_f)_R\times U(N_f)_L$ to the diagonal group 
$U(N_f)_V$.

In either case, the remaining non-tachyonic mode of the bi-fundamental 
field, which we will continue to call the ``tachyon" $T$, is crucial for describing
the quark mass and condensate.

In this paper we  incorporate the tachyon into the 8-brane action using a proposal
of Garousi for the brane-antibrane effective action \cite{Garousi:2004rd}, 
which extends Sen's original proposal for the non-BPS D-branes \cite{Sen:1999md}.
We  show that the coupled equations of motion for the tachyon $T$
and the 8-brane-anti-8-brane separation $L$ admit a solution 
which describes a U-shaped configuration.
In our solution the tachyon has a non-trivial profile, which for large $u$
is a linear combination of a normalizable mode and a non-normalizable mode.
We  relate the coefficient of the former to the quark condensate 
$\langle\bar{q}q\rangle$, 
and the coefficient of the latter to the quark mass $m_q$.
We  also show that the pions, which are part of the meson spectrum, 
acquire a mass that satisfies the GOR relation (\ref{GOR}).\footnote{A different
approach to the pion mass in this model was discussed in \cite{Hashimoto:2007fa}.} 
For $m_q=0$ our solution describes the same configuration
as \cite{Sakai:2004cn,Antonyan:2006vw}, but it also includes the 
effect of the (normalizable mode of the) massive bi-fundamental field.
At large $u$ the solutions are the same, but the precise shape of the 8-brane
at finite $u$ changes.

For simplicity we will consider the non-compact case dual to the NJL model
\cite{Antonyan:2006vw}, namely the 
near-horizon background of $N_c$ extremal D4-branes.
The metric in this case does not contain the ``thermal factor"
$f(u)=1-u_{KK}^3/u^3$. The behavior near the boundary at 
$u\rightarrow\infty$ will be similar to the compact case since $f(\infty)=1$,
and therefore our results for the quark mass and condensate, which are
determined by the behavior of $T$ near the boundary, will be the same.
This is also reasonable from the field theory point of view.
While the gauge sector of this model is very different from QCD,
and Kaluza-Klein states do not decouple,
the flavor sector, which is where chiral symmetry breaking and quark masses are seen,
is the same.
We will also deal only with the one flavor case $N_f=1$, for which the 8-brane
theory is Abelian.
Note that in this case the would-be broken symmetry is the anomalous $U(1)_A$.
At large $N_c$, however, the anomaly, and with it the mass of the would-be
Goldstone boson $\eta'$, is suppressed \cite{Witten:1979vv,Veneziano:1979ec}
(see however \cite{Bergman:2006xn} for a discussion of how it is suppressed in this model).
The GOR relation (\ref{GOR}) therefore holds also for the $\eta'$.
\footnote{There is an alternative large $N_c$ extension of one-flavor QCD,
in which the fermions transform in the anti-symmetric representation of the
gauge group \cite{Armoni:2003fb}. In that model the anomaly
is not suppressed, and the GOR relation is not expected to hold for the $\eta'$.}

A holographic dual description of the chiral condensate and quark
mass in QCD has been discussed previously in the context of the
``bottom-up" AdS/QCD model \cite{Erlich:2005qh,Da Rold:2005zs}, 
which is essentially a five-dimensional
$U(N_f)\times U(N_f)$ Yang-Mills theory in $AdS_5$ with a bi-fundamental
tachyonic scalar field. This was later generalized to a 
tachyonic DBI + CS theory in \cite{Casero:2007ae}.

\medskip

The outcome  of the present paper is a holographic picture  where 
\begin{itemize}
\item
 The spontaneous breaking of flavor chiral symmetry emerges from a 
 ``Higgs mechanism'' with an order parameter which is the expectation value of
the bi-fundamental tachyon field. 
\item
 The  current algebra  mass of the quarks is associated with a non-normalizable mode of the tachyon. 
 The  quark anti-quark condensate can be identified with a normalizable mode of the tachyon.
\item
The pions of the model obey the GOR relation.
\end{itemize}    

The paper is organized as follows.
In section 2 we review the proposal for the D$p$-$\overline{\mbox{D}p}$ 
effective action, and apply it to the D$8$-$\overline{\mbox{D}8}$ system
in the near-horizon extremal 4-brane background.
In section 3 we  study the asymptotic forms of the solutions for 
$T(u)$ and $L(u)$, both at large $u$ and near the point $u=u_0$ where the branes 
and anti-branes connect. We  extract the quark mass and condensate from
the behavior of $T$ at large $u$.
In section 4 we present numerical solutions which interpolate between the 
two asymptotic solutions, and compare with the solution without the tachyon of
\cite {Antonyan:2006vw}.
In section 5 we  begin to analyze the meson spectrum in the tachyon background.
This  includes both the fluctuations of the scalar fields $T$ and $L$,
as well as the worldvolume gauge fields on the 8-branes and anti-8-branes.
In particular we show that the mass of the pions satisifies the GOR relation.

\section{The D$8$-$\overline{\mbox{D}8}$ theory}

A proposal for the effective action of a parallel $p$-brane-anti-$p$-brane
system in curved spacetime was given by Garousi in \cite{Garousi:2004rd}.
Denoting by $X^{(n)}$ and $A^{(n)}$ the adjoint (position) scalar fields and gauge fields
on the branes ($n=1$) and anti-branes ($n=2$), and by $T$ the complex bi-fundamental 
scalar field,
the action is given by
\be
\label{Garousi_action}
S = -T_p \int d^{p+1}\,\sigma \sum_{n=1,2} e^{-\Phi(X^{(n)})}
V(T) \sqrt{1 + {|T|^2 |L|^2\over 2\pi\alpha'}} \sqrt{-\mbox{det}\, ({\cal G}^{(n)}
+{\cal T}^{(n)})} \,,
\ee
where $L \equiv X^{(1)}-X^{(2)}$ is the brane-antibrane separation, and
\be
{\cal G}_{ab}^{(n)} &=& P^{(n)}\left[G_{ab} - 
{|T|^2 \over 2\pi\alpha'\left(1+{|T|^2 |L|^2\over 2\pi\alpha'}\right)}
G_{ai} L^i L^j G_{jb} \right] + 2\pi\alpha' F^{(n)}_{ab} \\[5pt]
{\cal T}_{ab}^{(n)} &=& {1\over 1+{|T|^2 |L|^2\over 2\pi\alpha'}}
\bigg[
\pi\alpha' \left(D_aT(D_bT)^* + D_bT(D_aT)^*\right) \nonumber\\[5pt]
 && \mbox{} + {i\over 2} \left( G_{ai} + \partial_a X^{(n)j} G_{ji} \right) L^i 
 \left( T(D_bT)^* - T^* D_bT \right) \nonumber\\[5pt]
 && \mbox{} + {i\over 2}  \left( T(D_aT)^* - T^* D_aT \right) L^i 
 \left( G_{ib} - G_{ij} \partial_b X^{(n)j}  \right)
 \bigg] \,.
\ee
We use $a,b$ for the worldvolume directions, and $i,j$ for the transvese directions.
The covariant derivative of the bi-fundamental scalar is given by 
$D_a T = \partial_a T - i(A^{(1)}-A^{(2)}) T$,
and $V(T)$ is the scalar field potential.

This action was obtained by generalizing Sen's action for a non-BPS 
9-brane in Type IIA string theory \cite{Sen:1999md} as follows. 
First, the tachyon kinetic term is added 
under the square root \cite{Garousi:2000tr}.
Second, the action is extended to two unstable 9-branes
by the familiar symmetric trace prescription for the non-Abelian DBI action.
Third, the action is transformed to a 9-brane-anti-9-brane action
in Type IIB string theory by projecting with $(-1)^{F_L}$.
Finally, the general $p$-brane-anti-$p$-brane action is obtained by T-duality.
To separate the branes and antibranes we turn on a Wilson line
in the 9-brane-anti-9-brane model, which fixes the dependence on $L$.
As a check, note that for $L=0$ this reduces to Sen's action for a coinciding
brane and antibrane \cite{Sen:2003tm}.

\begin{figure}
\begin{center}
\includegraphics{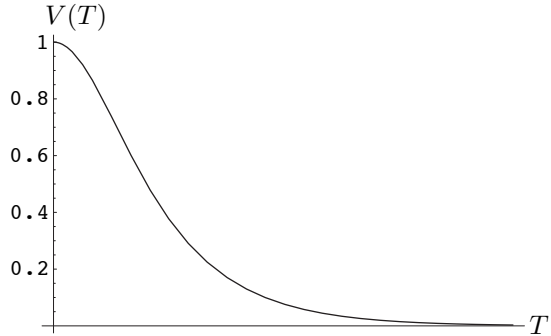}
\end{center}
\caption{The inverse cosh Tachyon potential.}
\label{tpot}
\end{figure}

The tachyon potential for the brane-antibrane pair is not known precisely
even in flat space.
Boundary superstring field theory gives a potential \cite{Kutasov:2000aq}
\be
\label{exp_potential}
 V(T) = e^{-T^2/4} \,.
\ee
An alternative proposal for the potential is 
\cite{Buchel:2002tj,Leblond:2003db,Lambert:2003zr,Garousi:2004rd}
\be
\label{cosh_potential}
 V(T) = {1\over \cosh(\sqrt{\pi} T)} \,.
\ee
This reproduces, for example, the S-brane solution using the tachyon effective theory
\cite{Lambert:2003zr}.
In both proposals (and there may be others) 
the true vacuum is at $T\rightarrow \pm \infty$, but the details are different.
We will work with the inverse cosh potential (\ref{cosh_potential}) shown 
in figure~\ref{tpot}.

\medskip

Let us apply this proposal to the D8-$\overline{\mbox{D}8}$ system in the non-compact
extremal D4-brane background. The background is defined by
\be
ds^2 = \left({U\over R}\right)^{3\over 2}\left(\eta_{\mu\nu}dx^\mu dx^\nu + (dx_4)^2\right)
+ \left({U\over R}\right)^{-{3\over 2}}\left(( dU)^2 + (U)^2 d\Omega_4^2\right) \,,
\ee
and
\be
e^\Phi = g_s \left({ U\over R}\right)^{3\over 4} \,,
\ee
where $\mu,\nu=0,1,2,3$, and $0\leq U < \infty$.
We assume that the 8-brane and anti-8-brane are positioned symmetrically at
$X_4^{(1)}=L/2$ and $X_4^{(2)}=-L/2$, respectively, 
and that the configuration depends only on the radial coordinate $U$.
It will also be convenient to work in the unitary gauge in which $T$ is real.
Suppressing the gauge fields for now,
the action for the D8-$\overline{\mbox{D}8}$ pair in this background
becomes
\be 
\label{the_action}
S[T,L] = \mbox{} - 2 {\cal N} \int d^4x du V(T) u^4 
\sqrt{{\cal D}[T,L]} \,,
\ee
where
\be
\label{D}
{\cal D}[T,L] = {1 \over u^3} + {1 \over 4R^2}(L'(u))^2 
	+{2\pi\alpha' \over R^2 u^{3 \over 2}}(T'(u))^2
	+ {1 \over 2\pi\alpha' u^{3 \over 2}}(L(u))^2(T(u))^2 \,,
\ee
and where we have defined $u\equiv U/R$ and 
${\cal N}\equiv \mu_8\Omega_4 R^5/g_s$.

Note that the proper distance between the 8-brane and anti-8-brane is 
\be
L_{proper} = u^{3/4} L \,,
\ee
so even if we keep the brane and antibrane well separated in coordinate distance,
the proper distance will decrease below the string scale for small enough $u$, 
and the field $T$ will be tachyonic in that region.
One can see this directly by expanding the action for small $T$,
which gives (after properly normalizing to get a canonical kinetic term)
\be
\label{tachyon_mass}
m^2_T(u) = \mbox{}  -{1\over 2\alpha'} + {(L_{proper}(u))^2\over (2\pi\alpha')^2} \,.
\ee
We recognize the first term as the zero-point energy of the open superstring in 
the NS sector in flat space, and the second term as the contribution of the proper
length of the open string. This result is most likely not precise.
First, the flat space result for the zero-point energy probably changes
in this background. We do not know how to compute it, since this is an RR
background. Second, the straight string
stretched between the 8-brane and anti-8-brane is not the minimal length
(and mass) string, and it actually prefers to curve down in
$u$ \cite{Antonyan:2006vw}.
However we believe that the qualitative result is still correct, namely
that $m^2_T<0$ below some critical $u$.
In other words the field $T$ has a localized tachyonic mode in a small region
near $u=0$. This is similar to the tachyon which appears at the intersection
of branes which meet at a small angle.
We therefore expect the ground state to correspond 
to the connected 8-brane configuration.

\subsection{The compact case}

In the Sakai-Sugimoto model $x_4$ is compact and the near-horizon metric is given by
\be
\label{confined_background}
ds^2 = \left({U\over R}\right)^{3\over 2} \left(\eta_{\mu\nu}dx^\mu dx^\nu
+ f(U)(dx_4)^2\right)
+ \left({U\over R}\right)^{-{3\over 2}}\left({(dU)^2\over f(U)}
+ U^2 d\Omega_4^2\right) \,,
\ee
where
\be
f(U) = 1 - {U^3_{KK}\over U^3} \;\; , \;\;
U_{KK} = {4\over 9}\, {R^3\over R_4^2} \,.
\ee
Strictly speaking, the 8-brane and anti-8-brane cannot be treated as separate 
entities in this background, since the $x_4$ circle shrinks to zero size at $U=U_{KK}$.
In this case the brane and anti-brane are necessarily connected,
and the D8-$\overline{\mbox{D}8}$ action should be viewed as a large-$u$ 
effective theory for the worldvolume fields on the two sides of the 8-brane,
together with the massive scalar field coming from the open string stretched between
the two sides. At large $U$ the compact background is essentially identical
to the non-compact one, so the results related to the mass and condensate will
be the same. The precise profiles of the fields $T(u)$ and $L(u)$ at finite $u$ will be different.

\section{Asymptotic solutions}

The equations of motion that follow from (\ref{the_action}) are given by
\be
\label{L_equation}
{d \over du}\left[
{V(T) \over \sqrt{\cal D}}{u^4 \over 4R^2} L' \right]&=& 
{V(T) \over \sqrt{\cal D}}u^{5 \over 2} L T^2 \\[5pt]
\label{T_equation}
{d \over du}\left[{V(T) \over \sqrt{\cal D}}{u^{{5 \over 2}} \over R^2} T' \right]
&=& {V(T) \over \sqrt{\cal D}}u^{{5 \over 2}} L^2 T + {dV(T) \over dT}u^4 \sqrt{\cal D} \,,
\ee
where ${\cal D}$ was defined in (\ref{D}), 
$V(T)$ is the inverse cosh potential, and we have set $2\pi\alpha' =1$.

The tachyon equation (\ref{T_equation}) has a trivial solution $T=0$.
In this case the solution to the $L$ equation (\ref{L_equation}) is
$L(u)=L_\infty$, corresponding to the parallel D8-$\overline{\mbox{D}8}$
configuration.\footnote{Equation (\ref{L_equation}) reduces in this
case to the same equation one gets from the single 8-brane action 
without the tachyon \cite{Antonyan:2006vw}.
There are two solutions in that case corresponding to a straight 8-brane
and a U-shaped 8-brane.
However the action in our case is doubled since it includes
both an 8-brane and an anti-8-brane.
Consequently there are four possible solutions with $T=0$,
corresponding to either brane or antibrane being straight or U-shaped.
We are interested only in solutions with two asymptotic boundaries, one for the
8-brane and one for the anti-8-brane (or equivalently we require $X_4^{(1)}(u)$
and $X_4^{(2)}(u)$ to be single-valued). With $T=0$ that leaves only the 
straight and parallel  D8-$\overline{\mbox{D}8}$ solution.}
This configuration is unstable due to the localized tachyon mode near $u=0$.
The stable solution must involve a non-trivial tachyon condensate $T(u)$,
which, as we shall see below, corresponds to a single U-shaped configuration.

We expect the 8-brane and anti-8-brane to connect roughly at the radial position
below which the bi-fundamental field is tachyonic.
Let us first expand the fields near this point:
\be
L(u) &=& (u-u_0)^p[l_0 + l_1 (u-u_0) +\cdots] \\[5pt]
T(u) &= & (u-u_0)^q[t_0 + t_1 (u-u_0) +\cdots]  \,,
\ee
where we assume that $l_0,t_0>0$.
To leading order, the $L$ equation (\ref{L_equation}) gives
\be
q=-2 \;\;\; \mbox{and} \;\;\; t_0 = {\sqrt{\pi} \over 2R^2}\, p\, u_0^{3 / 2} \,.
\ee
This implies, in particular, that $p>0$. 
In addition, the absence of sources at $u=u_0$ implies that $p<1$,
so that $L'(u_0)\rightarrow\infty$ and the configuration is smooth.
The $T$ equation (\ref{T_equation}) is then also satisfied
to leading order.
The leading behavior near $u_0$ is then
\be
\label{IR}
T(u)\sim (u-u_0)^{-2} \;\; , \;\;
L(u) & \sim (u-u_0)^p \;\; , \;\; 0<p<1 \,.
\ee
This is in accord with the interpretation of the non-trivial solution as the
chiral-symmetry-breaking U-shaped configuration. 
The brane-antibrane separation vanishes at $u=u_0$, and the tachyon diverges, 
{\em i.e.} goes to its true vacuum in the potential (\ref{cosh_potential}). 

To compute the gauge theory quantities, in this case the quark mass and condensate,
we should look at the behavior of the solution at large $u$.
This corresponds to the UV limit of the gauge theory.
Strictly speaking, the UV limit is not well-defined in this model, since it is really
a five-dimensional gauge theory. We will therefore always be considering 
a UV cutoff $u_\infty$.
In this regime the field $T$ is very massive, so we can consider small fluctuations
away from the trivial solution:
\be
L(u) &=& L_\infty + \tilde{L}(u) \\[5pt]
T(u) &=& 0 + \tilde{T}(u) \,,
\ee 
where $\tilde{L}\ll L_\infty$ and $\tilde{T}\ll 1$.
In this approximation the action is quadratic
\be
\label{asymptotic_action}
S \propto \int d^4x du\, \left[
u^{5/2} + {1\over 8R^2} u^{11/2} (\tilde{L}')^2 + 
u^4\left({1\over 2R^2} (\tilde{T}')^2 + {L^2_\infty\over 2} (\tilde{T})^2\right) \right] \,.
\ee
The asymptotic solutions for $u\gg 1$ are given by
\be
\label{L_UV}
\tilde{L}(u) &\approx & C_L u^{-9/2} \\[5pt]
\label{T_UV}
\tilde{T}(u) & \approx & u^{-2}\left(C_Te^{-RL_\infty u} + C_T' e^{+RL_\infty u}\right) \,.
\ee
These solutions are only valid in a regime of $u$ for which the
fluctuations are small.
We therefore have to assume that 
$C'_T\lesssim e^{-RL_\infty u_\infty}$,
whereas $C_T$ and $C_L$ can be taken to be ${\cal O}(1)$ in the cutoff.


\subsection{Quark mass and condensate}

The growing and decaying exponentials correspond to the non-normalizable and normalizable
solutions for $\tilde{T}$, respectively.
We would therefore like to identify the coefficients $C_T'$ and 
$C_T$ with the
quark mass $m_q$ and quark condensate $\langle \bar{q} q \rangle$,
respectively. 
Let us verify this explicitly. In QCD (at zero temperature) the quark condensate 
is given by the variation of the energy density with respect to the quark mass
\be
\label{quark_condensate}
\langle \bar{q} q \rangle = \left.{\delta {\cal E}_{QCD} \over \delta m_q} 
\right|_{m_q=0}\,.
\ee
Let us assume that $m_q$ is given by the (dimensionless) parameter $C_T'$
\be
m_q = \Lambda C_T' \,, 
\label{quark_mass}
\ee
where $\Lambda$ is some fixed mass scale. 
To evaluate (\ref{quark_condensate}) in the holographic dual we must vary 
the asymptotic (Euclidean) 8-brane action (\ref{asymptotic_action}) with respect to the 
parameter $C_T'$ of the solution.
The general variation is
\be
\delta S = \int du \left[{\delta{\cal L}\over\delta T(u)}\delta T(u)
+ {\delta{\cal L}\over\delta T'(u)} \delta T'(u)
+ {\delta{\cal L}\over\delta L(u)} \delta L(u)
+ {\delta{\cal L}\over\delta L'(u)} \delta L'(u)
\right] \,.
\ee
Using the equations of motion this reduces to
\be
\delta S = \left.{\delta{\cal L}\over\delta T'(u)} \delta T(u) \right|_{u_0}^\infty
+  \left.{\delta{\cal L}\over\delta L'(u)} \delta L(u) \right|_{u_0}^\infty \,.
\ee
Focussing on the variation with respect to the tachyon we find
\be
\delta S = \mbox{} - \left. {2{\cal N} \over R^2} \,
   {u^{5/2} V(T) T'(u)\over\sqrt{\cal D}} \delta T(u)\right|_{u_0}^\infty \,.
\ee
Only the upper limit contributes, since although both $T$ and $T'$ diverge
in the lower limit, the potential $V(T)\sim \exp(-\sqrt{\pi}T)\rightarrow 0$
much faster.
Using the large $u$ asymptotic form of the solution (\ref{T_UV}) we find
for a variation with respect to $C_T'$:
\be
{\delta S\over\delta C'_T} = -{2{\cal N} \over R} L_\infty \left( C_T - C'_T e^{2RL_\infty u_\infty}\right) \,,
\ee
where we have imposed the cutoff $u_\infty$.
Since $C_T'$ is identified with the quark mass, we find that 
the quark condensate is related to $C_T$ as
\be
\langle \bar{q} q \rangle = {2{\cal N}L_\infty\over \Lambda R}\, C_T \,. 
\label{C_T-qq}
\ee

\section{Numerical solutions}

The asymptotic solutions near $u=u_0$ and at large $u$ must connect 
in the full solution to the equations of motion 
(\ref{L_equation}) and (\ref{T_equation}). 
In this section we present a numerical analysis of these equations.
For convenience we define the dimensionless quantities (recall that $2\pi\alpha' =1$)
\be
y\equiv 4R^4 u ,\quad f(y) \equiv {1 \over 4R^3}L(u) ,\quad g(y) \equiv \sqrt{2}\, T(u)\,. 
\ee
In terms of these the D8-$\overline{\mbox{D}8}$ action (\ref{the_action}) becomes
\be
\label{num_act}
S &=& -{{\cal N} \over 64R^{14}}\int d^4x dy 
{y^4 \sqrt{\tilde {\cal D}}\over \cosh\bigl(\sqrt{\pi \over 2}g(y)\bigr)} \,,
\ee
where 
\be
{\tilde {\cal D}} = y^{-3} +f'(y)^2 +y^{-3/2}g'(y)^2 +y^{-3/2}f(y)^2g(y)^2 \,,
\ee
and the equations of motion become
\be
\label{numL_equation}
&&{d \over dy}\biggl[{y^4 {\tilde {\cal D}}^{-1/2} f'(y) \over \cosh\bigl(\sqrt{\pi \over 2}g(y)\bigr)}\biggr] 
	= {y^{5/2} {\tilde {\cal D}}^{-1/2} f(y)g(y)^2 \over \cosh\bigl(\sqrt{\pi \over 2}g(y)\bigr)} 
	\\[5pt]
\label{numT_equation}
&&{d \over dy}\biggl[{y^{5/2} {\tilde {\cal D}}^{-1/2} g'(y) \over \cosh\bigl(\sqrt{\pi \over 2}g(y)\bigr)}\biggr] 
	= {y^{5/2} {\tilde {\cal D}}^{-1/2} f(y)^2g(y) \over \cosh\bigl(\sqrt{\pi \over 2}g(y)\bigr)} 
	-\sqrt{\pi \over 2}{\tanh\bigl(\sqrt{\pi \over 2}g(y)\bigr) \over \cosh\bigl(\sqrt{\pi \over 2}g(y)\bigr)}y^4{\tilde {\cal D}}^{1 \over 2} \,.  
\ee
The range for $y \in [0,\infty)$ will be approximated numerically by the range $[0.01,100]$.

\begin{figure}
\begin{center}
\includegraphics{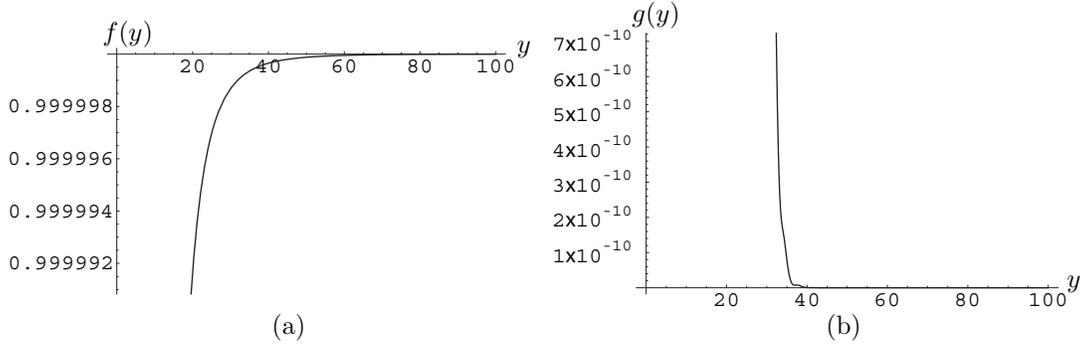}
\end{center}
\caption{(a) The D8-$\overline{\mbox{D}8}$ separation. (b) The tachyon profile. }
\label{figuv}
\end{figure}

The solution is fixed by imposing boundary conditions 
for $f(y)$, $g(y)$, and their derivatives, either at infinity (UV) or at $y=1$ (IR),
which corresponds roughly to $u_0$.
Let's look at UV boundary conditions first.
Guided by the UV asymptotic form of the solution 
(\ref{L_UV}) and (\ref{T_UV}), we impose
\be
\label{ini_uv}
f(100) = 1 ,\quad 
f'(100) = 10^{-10} ,\quad
g(100) = 10^{-30} ,\quad
g'(100) = -10^{-30} .
\ee
Figure \ref{figuv}(a) shows the resulting numerical solution for the shape
of the 8-brane, and figure \ref{figuv}(b) shows the tachyon profile for this
solution. The tachyon increases as $y$ decreases, and blows up,
{\em i.e.} attains its vacuum value,
where the 8-brane and anti-8-brane connect.

\begin{figure}
\begin{center}
\includegraphics{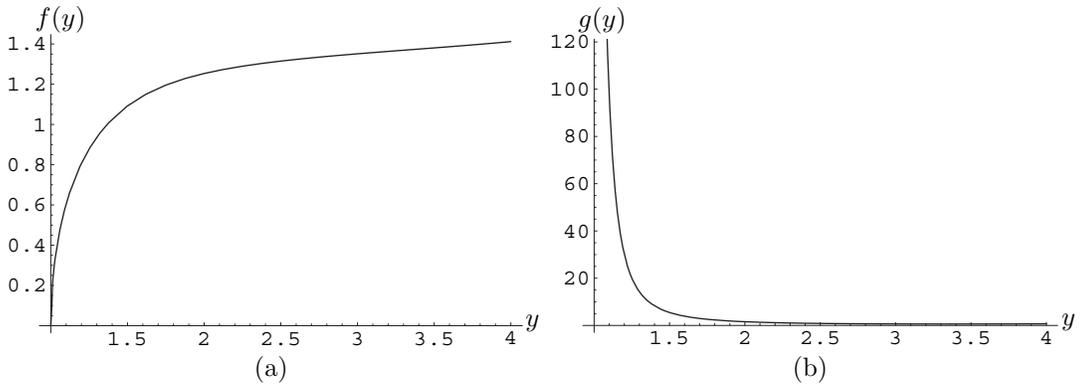}
\end{center}
\caption{(a) The shape of the 8-brane, and (b) the profile of the tachyon from
the IR.}
\label{figir}
\end{figure}

\begin{figure}
\begin{center}
\includegraphics{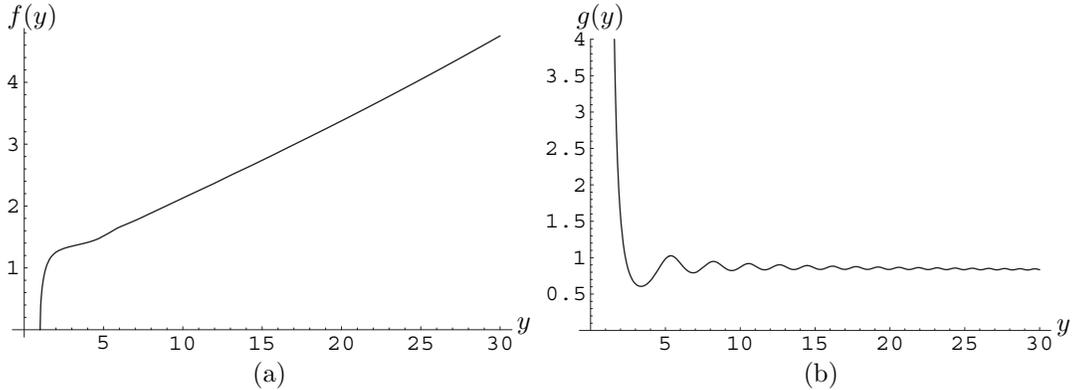}
\end{center}
\caption{Same as figure 3, but going to larger $y$.}
\label{figirlarge}
\end{figure}

Now let's look at IR boundary conditions. 
Guided by the IR asymptotics (\ref{IR})
we impose the numerical boundary conditions
\be
\label{IR_BC}
f(1) = 0.001, \quad
f'(1) = 500, \quad
g(1) = 400, \quad
g'(1) = -16000 \,. 
\ee
The solution near the connection point at $y=1$ is shown in
figure \ref{figir}.
The behavior is qualitatively the same as with the UV boundary conditions.
When we look at larger $y$, however, the qualitative behavior changes
(figure \ref{figirlarge}). This is due to the sensitivity of the numerical solution
to the IR boundary conditions (\ref{IR_BC}).
The exact IR boundary values of $f',g$ and $g'$ are infinite.
As we increase the numerical IR boundary values of $f',g$ and $g'$
the region where the behavior changes moves to larger and larger $y$.

\subsection{A comparison with the AHJK solution}

\begin{figure}
\begin{center}
\includegraphics{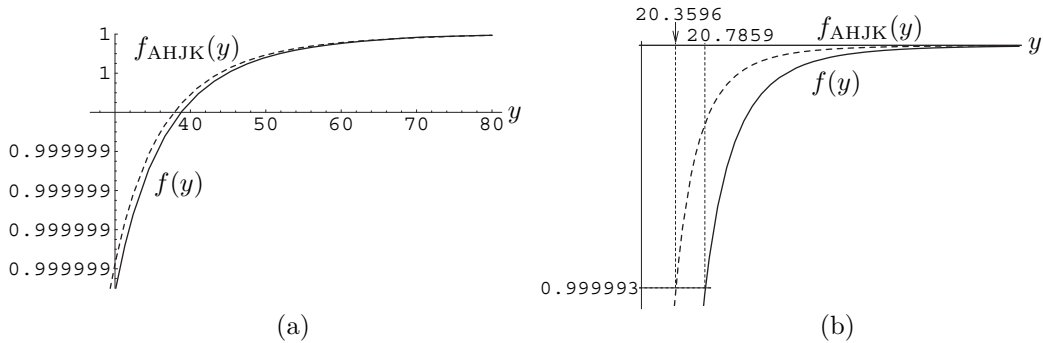}
\end{center}
\caption{The solid line describes $f(y)$ in our tachyonic model. The dashed line describes the AHJK solution $f_{\rm AHJK}(y)$.}
\label{figahjk}
\end{figure}

Let us compare this solution with the solution for the single 8-brane action
without the tachyon found in \cite{Antonyan:2006vw}. 
The latter is just the solution of (\ref{numL_equation}) with $g(y)=0$:
 \be
{d \over dy}\left[{y^4 f_{\rm AHJK}'(y) \over \sqrt{y^{-3} + f_{\rm AHJK}'(y)^2}}\right] = 0\,. 
\ee
Using the same UV boundary conditions we arrive at the numerical solution presented
in figure \ref{figahjk}, where we also present our solution for comparison.
The two configurations are very close in this regime, which is understandable since the correction due to the very massive field $T$ is small.

We would like to argue that our solution with the non-trivial tachyon profile
is, in some sense, a better approximation to the exact string theory solution.
As evidence for this we will show that the free energy of our solution is smaller
than that of the solution of \cite{Antonyan:2006vw}, which we will refer to
as the AHJK solution.\footnote{Recall also that the U-shaped solution of 
\cite{Antonyan:2006vw}
has a lower free energy than that of the parallel configuration.}
The free energy is given by the Euclidean action of the solution, which
for our solution is
\be 
{\cal E}[f(y),g(y)] = {{\cal N} \over 64R^{14}}\int dy 
	{y^4 \sqrt{y^{-3} +f'(y)^2 +y^{-3/2}g'(y)^2 +y^{-3/2}f(y)^2g(y)^2} 
	\over \cosh\bigl(\sqrt{\pi \over 2}g(y)\bigr)} \,.
\ee 
The free energy for the AHJK solution is  
given by  
\be 
{\cal E}_{\rm AHJK}[f_{\rm AHJK}(y)] = {{\cal N} \over 64R^{14}}\int dy\, y^4\sqrt{y^{-3}+f_{\rm AHJK}'(y)^2} \,.  
\ee  
By splitting the integration over $y$ to a UV part and an IR part,  
we can estimate the free energy of our tachyonic solution  
and the AHJK solution.  
In the IR region, the free energy of the tachyonic solution  
${\cal E}^{\rm IR}$ is strongly suppressed by the exponentially  
vanishing factor $\cosh^{-1}(\sqrt{\pi/2}g(y))$,  
which comes from the tachyon potential.  
So ${\cal E}^{\rm IR}$ is obviously much smaller  
than the IR part of the AHJK free energy  
${\cal E}^{\rm IR}_{\rm AHJK}$.  
On the other hand, in the UV region, we shall compare ${\cal E}$ and  
${\cal E}_{\rm AHJK}$ numerically. From figure~\ref{figahjk}(b),
and using the numerical solutions of $f,g,f_{\rm AHJK}$, we calculate\footnote{ 
The lower bounds of the integration intervals depend on the capacity of our computer.}  
\be 
&&{64R^{14} \over {\cal N}}({\cal E}^{\rm UV} - {\cal E}^{\rm UV}_{\rm AHJK}) \nonumber \\ 
&=& \int_{20.7859}^{100}dy {y^4\sqrt{y^{-3}+f'^2+y^{-3/2}g'^2+y^{-3/2}f^2g^2} \over \cosh\bigl(\sqrt{\pi \over 2}g\bigr)} 
	- \int_{20.3596}^{100} dy\, y^4\sqrt{y^{-3}+{f_{\rm AHJK}'}^2} \nonumber \\ 
&=&-818.417 < 0,  
\ee 
that is to say, ${\cal E}^{\rm UV} < {\cal E}^{\rm UV}_{\rm AHJK}$.  
Combining the results in the IR and UV regions, we see that 
\be 
{\cal E} < {\cal E}_{\rm AHJK} \,,  
\ee 
so our tachyonic solution appears to be more favorable than  
the AHJK solution.

\section{The meson spectrum}

We now turn to the analysis of the spectrum in the tachyon U-shaped background.
The fluctuations of $T,L$ and the gauge fields on the 8-brane and
anti-8-brane correspond to various mesons, including scalars,
pseudo-scalars, vectors and axial-vectors.
We are interested mainly in the lowest pseudo-scalar modes (the pions).
In particular, we would like to see how they acquire mass 
when the quarks are massive.
For completeness we also set up the eigenvalue problems for the other mesons, 
but we leave the (numerical) analysis for future work.

\subsection{Scalar fields}

Let's start with the scalar fields $L$ and $T$.
We expand around the classical solution
\be
T(x^\mu,u) &=& T(u) + t(x^\mu,u) \\
L(x^\mu,u) &=& L(u) + \ell(x^\mu,u) \,,
\ee
where $t(x^\mu,u)$ and  $\ell(x^\mu,u)$ are real scalar
fields\footnote{In general there is also
a pseudo-scalar fluctuation of the phase of the tachyon $\theta(x^\mu,u)$, 
but we are working in unitary gauge where $\theta=0$.}. Expanding the 
8-brane-anti-8-brane action to quadratic order gives
\be
\label{tl_action}
S[t,\ell] &=& -{\cal N}\int d^4x du\, \bigg[
{\cal I}_1 (\partial_\mu t)^2 +{\cal I}_2 (\partial_\mu \ell)^2 +{\cal I}_3  \partial_\mu t \partial^\mu \ell
	+{\cal I}_4 (t')^2 + {\cal I}_5 t^2 +{\cal I}_6 tt' \nonumber \\
&&	+{\cal I}_7 (\ell')^2 + {\cal I}_8 \ell^2 +{\cal I}_9 \ell\ell'
	+{\cal I}_{10} t'\ell' + {\cal I}_{11} t'\ell +{\cal I}_{12} t\ell' +{\cal I}_{13} t\ell \bigg] \,,
\ee
where the coefficients are given by
\be	
& & {\cal I}_1 := {V(T) \over \sqrt{\cal D}}\left(u^{-{1 \over 2}}+{u^{5 \over 2} (L')^2 \over 4R^2(1 + u^{3 \over 2}L^2 T^2)}\right) , \quad
	{\cal I}_2 := {V(T) \over \sqrt{\cal D}}\left({u \over 4}+{u^{5 \over 2} (T')^2 \over 4R^2(1 + u^{3 \over 2}L^2 T^2)}\right) , \nonumber \\
& & {\cal I}_3 := -{V(T) \over \sqrt{\cal D}}{u^{5 \over 2}L'T' \over 2R^2(1 + u^{3 \over 2}L^2 T^2)} , \quad
	{\cal I}_4 := {V(T) \over \sqrt{\cal D}}\left({u^{5 \over 2} \over R^2}-{u (T')^2 \over R^4 {\cal D}}\right) , \nonumber \\
& &{\cal I}_5 := -{V(T) u L^4 T^2 \over {\cal D}^{3 \over 2}} 
		+ {V(T) u^{5 \over 2} L^2\over \sqrt{\cal D}}
		+{dV(T) \over dT}{2u^{5 \over 2}L^2 T \over \sqrt{\cal D}}
		+{d^2V(T) \over dT^2}u^4\sqrt{\cal D} , \nonumber \\
& &{\cal I}_6 := -{2V(T) u L^2 T T' \over R^2 {\cal D}^{3 \over 2}}
		+{dV(T) \over dT}{2u^{5 \over 2}T' \over R^2\sqrt{\cal D}} , \quad
	{\cal I}_7 := {V(T) u^4 \over 4R^2 \sqrt{\cal D}} 
		-{V(T) u^4 (L')^2 \over 16 R^4 {\cal D}^{3 \over 2}} , \nonumber \\
& &{\cal I}_8 := {V(T) u^{5 \over 2}T^2 \over \sqrt{\cal D}} 
		-{V(T)uL^2T^4 \over {\cal D}^{3 \over 2}} , \quad
	{\cal I}_9 := -{V(T)u^{5 \over 2}LL'T^2 \over 2R^2{\cal D}^{3 \over 2}} , \quad 
	{\cal I}_{10} := -{V(T)u^{5 \over 2}L'T' \over 2 R^4 {\cal D}^{3 \over 2}} , \nonumber \\ 
& &{\cal I}_{11} := -{2V(T)uLT^2T' \over R^2 {\cal D}^{3 \over 2}} , \quad 
	{\cal I}_{12} := -{V(T)u^{5 \over 2}L^2L'T \over 2R^2{\cal D}^{3 \over 2}} 
	+{dV(T) \over dT}{u^4 L' \over 2R^2\sqrt{\cal D}}	, \nonumber \\
& &{\cal I}_{13} := -{2V(T)u L^3T^3 \over R {\cal D}^{3 \over 2}}
		+{4V(T)u^{5 \over 2}LT \over \sqrt{\cal D}} 
		+{dV(T) \over dT}{2 u^{5 \over 2}LT^2 \over \sqrt{\cal D}} . \nonumber 
\ee
The four-dimensional mass matrix will get contributions from all the quadratic terms
in $t,t',\ell,\ell'$.

\subsection{Gauge fields}\label{Sec:gauge_field}

Now consider the gauge fields $A^{(1)}$ and $A^{(2)}$.
We will use the symmetric and anti-symmetric combinations
\be
A^{\pm}(x^\mu,u) \equiv {1\over 2}(A^{(1)}(x^\mu,u) \pm A^{(2)}(x^\mu,u)) \,,
\ee
and consider only the $\mu$ and $u$ components.
The former will give rise to vector ($A^{+}_\mu$) and axial-vector 
($A^{-}_\mu$) mesons, and the latter to scalar and pseudo-scalar mesons
in four dimensions.
Let us also further fix the gauge by setting
\be
A^+_u = 0 \,.
\ee
Expanding the action to quadratic order in the gauge fields then
gives
\be
\label{gauge_action}
S[A^+,A^-] &=& \mbox{} - {\cal N} \int d^4x du\, u V(T) \sqrt{\cal D}
\biggl[
 {1 \over 2} |F^{+}_{\mu\nu}|^2 + {1 \over 2}|F^{-}_{\mu\nu}|^2 
 + {1 + u^{3\over 2} T^2 L^2\over R^2 {\cal D}}\left(	
	|A^{+\prime}_\mu|^2 + |F^{-}_{\mu u}|^2
	\right) \nonumber\\
& + & {4u^{3\over 2} T^2\over 1 + u^{3\over 2}T^2 L^2}
\left(1 + {T^2 L^2 (L')^2\over 4u^{3\over 4}R^2 {\cal D}}\right) (A^{-}_\mu)^2 
 + {4 u^{3\over 2} T^2\over R^2{\cal D}} (A^{-}_u)^2 
 + {2 u^{3\over 2} T^2 L L'  \over  R^2 {\cal D}}F^{-}_{\mu u} A^{-\mu} \biggr] \,.
 \nonumber \\
\ee

\subsubsection{The $A^+$ sector}

The action in the symmetric (vector) sector is given by
\be
S[A^+] = -{\cal N}\int d^4x du\, u V(T) \sqrt{\cal D}\left[
{1 \over 2} |F^{+}_{\mu\nu}|^2 
+ {1 + u^{3\over 2} T^2 L^2\over R^2 {\cal D}}|A^{+\prime}_\mu|^2
\right] . 
\ee
This sector is similar to the gauge field in the single 8-brane case in 
\cite{Sakai:2004cn}. We expand the gauge field $A^+_\mu$ in radial
modes $\psi_n(u)$
\be
A^+_\mu(x^\mu,u) = \sum_n a^{+(n)}_\mu(x^\mu)\psi_n(u) \,,
\ee
that satsify the eigenvalue equation
\be
\label{psi_eigen}
-{1 \over V(T) u \sqrt{\cal D}}\partial_u\left(V(T){u+u^{5 \over 2}T^2L^2 \over R^2 \sqrt{\cal D}}\psi_n'\right) 
	= (m_n^{+})^2\psi_n \,, 
\ee
and the normalization condition
\be
\label{psi_norm}
{\cal N}\int du\, u V(T) \sqrt{\cal D}\psi_m\psi_n = {1 \over 2} \delta_{mn} \,.
\ee
The four-dimensional action in this sector is then
\be
S[a_\mu^{+(n)}] = - \int d^4x \sum_{n=1}^\infty \left(
	{1 \over 4}f^{+(n)}_{\mu\nu}f^{+(n)\mu\nu} 
	+{1 \over 2}(m_n^{+})^2 a^{+(n)}_\mu a^{+(n)\mu}
\right) , 
\ee
where $f^{+(n)}_{\mu\nu} \equiv \partial_\mu a^{+(n)}_\nu - \partial_\nu a^{+(n)}_\mu$. 

The zero mode $\psi_0$, with $m_0^{+}=0$, is special.
The eigenvalue equation (\ref{psi_eigen}) gives
\be
\label{psi_zero}
\psi_0(u) \sim \int^u dv {R^2 \sqrt{\cal D} \over V(T(v))(v+v^{5 \over 2}T^2L^2)} \,. 
\ee
In the UV asymptotic region this becomes
\be
\psi_0 \sim \int_\infty^u dv\, v^{-21/2} \,,
\ee
so $\psi_0(u)$ is UV-normalizable. 
On the other hand, in the IR asymptotic region we get
\be
\psi_0(u) \sim \int_{u_0}^u dv 
\exp{\sqrt{\pi}t_0 \over (v-u_0)^2} \times 
	\begin{cases}
	v^{-7/4}(v-u_0)^{-3}  &(p>2) \\
	(v^{-7/4}+\ell_0^2t_0^2v^{-1/4})(v-u_0)^{-3}  &(p=2) \\
	\ell_0^2t_0^2v^{-1/4}(v-u_0)^{2p-7} \,. &(p<2)  
	\end{cases}
\ee
The exponential divergence of the integrand as
$v \to u_0$ implies that $\psi_0(u)$ is non-normalizable in the IR.
We therefore have to exclude this mode from the spectrum.
The spectrum of vectors is therefore purely massive.
Note that the exponential divergence is due
to the tachyon.

\subsubsection{The $A^-$ sector}

The action in the anti-symmetric sector is given by
\be
\label{A-_action}
S[A^-] = \mbox{} - {\cal N} \int d^4x du\, u V(T) \sqrt{\cal D}
\left[
{1 \over 2}|F^{-}_{\mu\nu}|^2 
 + {\cal B}_1 |F^{-}_{\mu u}|^2 
 + {\cal B}_2 |A^{-}_\mu|^2 
 + {\cal B}_3 (A^{-}_u)^2 
 + {\cal B}_4 F^{-}_{\mu u} A^{-\mu} 
\right] ,\nonumber\\
\ee
where the coefficients are given by
\be
\label{Bcoefficients}
{\cal B}_1 = {1 + u^{3\over 2} T^2 L^2\over R^2 {\cal D}} ,\,
{\cal B}_2 = {4u^{3\over 2} T^2\over 1 + u^{3\over 2}T^2 L^2}  
\left(1 + {T^2 L^2 (L')^2 \over 4u^{3\over 4}R^2 {\cal D}}\right) ,\,
{\cal B}_3 = {4 u^{3\over 2} T^2\over R^2{\cal D}} ,\,
{\cal B}_4 = {2 u^{3\over 2} T^2 L L'  \over  R^2 {\cal D}} \,. \nonumber\\
\ee
We decompose the four-dimensional part of the gauge field $A^{-}_\mu$ into 
a longitudinal component $A^\parallel_\mu$  and transverse components 
$A^\perp_\mu$ (where $\partial_\mu A^{\perp \mu}=0$), and expand all
the components in radial modes
\be
A^\perp_\mu(x^\mu,u) &=& \sum_n a^{-(n)}_\mu(x^\mu)\xi^\perp_n(u) , \\
A^\parallel_\mu(x^\mu,u) &=& \sum_n 
\partial_\mu\omega^{(n)}(x^\mu)\xi^\parallel_n(u) , \\
A^{-}_u(x^\mu,u) &=& \sum_n \omega^{(n)}(x^\mu)\zeta_n(u) \,.
\ee
These modes satisfy the eigenvalue equations
\be
\partial_u\left[uV(T)\sqrt{\cal D}
	\left({\cal B}_1 \xi^{\perp\prime}_n -{1 \over 2}{\cal B}_4\xi^\perp_n\right)\right] 
&=& uV(T)\sqrt{\cal D}\left({\cal B}_2 \xi^\perp_n 
- {1 \over 2}{\cal B}_4 \xi^{\perp\prime}_n - (m^{-}_n)^2 \xi^\perp_n\right) 
\label{eigen_xiperp} \\
{\cal B}_3 \zeta_n &=& M_n^2 \left[{\cal B}_1(\zeta_n - \xi^{\parallel\prime}_n)
	+{1 \over 2}{\cal B}_4\xi^\parallel_n\right] \,, 
\label{eigen_zeta} 
\ee
and an additional equation relating the longitudinal and pseudo-scalar modes
\be
-\partial_u\left[
	uV(T)\sqrt{\cal D}\left({\cal B}_1 (\zeta_n- \xi^{\parallel\prime}_n)
	+{1 \over 2}{\cal B}_4 \xi^\parallel_n\right)\right] 
=uV(T)\sqrt{\cal D}\left({\cal B}_2\xi^\parallel_n
	+{1 \over 2}{\cal B}_4(\zeta_n- \xi^{\parallel\prime}_n) \right) ,
\label{eigen_zetaxi} 
\ee
and the normalization conditions are given by
\be
{\cal N}\int_{u_0}^\infty du\, uV(T)\sqrt{\cal D}\xi^\perp_m\xi^\perp_n 
= {1 \over 2}\delta_{mn}  \,,
\label{norm_xiperp}\\
{\cal N}\int_{u_0}^\infty du\, uV(T)\sqrt{\cal D}\biggl[
{\cal B}_1 (\zeta_m-\xi^{\parallel\prime}_m)(\zeta_n-\xi^{\parallel\prime}_n) 
	+{\cal B}_2\xi^\parallel_m\xi^\parallel_n \nonumber\\
+{1 \over 2}{\cal B}_4 \bigl\{(\zeta_m-\xi^{\parallel\prime}_m)\xi^\parallel_n 
+ \xi^\parallel_m(\zeta_n-\xi^{\parallel\prime}_n)\bigr\}
\biggr] = {1 \over 2}\delta_{mn}  . 
\label{norm_xipara}
\ee
The four dimensional action in this sector then becomes
\be
\label{finalA-_act}
S[a^{-(n)}_\mu,\omega^{(n)}] &=& -{1 \over 2}\int d^4x \biggl[\sum_n
\biggl({1 \over 2} f^{-(n)}_{\mu\nu}f^{-(n)\mu\nu}
	+ (m^{-}_n)^2 a^{-(n)}_\mu a^{-(n)\mu}\biggr) \cr
	&&+ \sum_n\Bigl(\partial_\mu\omega^{(n)}\partial^\mu\omega^{(n)}
	+ M_n^2(\omega^{(n)})^2\Bigr)\biggr] ,
\ee
where $f^{-(n)}_{\mu\nu} \equiv \partial_\mu a^{-(n)}_\nu - \partial_\nu a^{-(n)}_\mu$. 

Using (\ref{eigen_xiperp}), (\ref{eigen_zeta}),
(\ref{eigen_zetaxi}), 
(\ref{norm_xiperp}) and (\ref{norm_xipara}), 
we can express the masses $m^{-}_n$ and $M_n$ as
\be
(m^{-}_n)^2 &=& 2{\cal N}\int du\, uV(T)\sqrt{\cal D}\left[
{\cal B}_1 (\partial_u\xi^\perp_n)^2 
	+{\cal B}_2(\xi^\perp_n)^2
	-{\cal B}_4 \xi^\perp_n\partial_u\xi^\perp_n
\right] , \label{mass_axialvec}\\
(M_n)^2 &=& 2{\cal N}\int du\, uV(T)\sqrt{\cal D} {\cal B}_3 (\zeta_n)^2. \label{mass_pscalar}
\ee
The spectrum in this sector consists of massive axial-vectors 
$a^{-(n)}_\mu$ and massive pseudo-scalars $\omega^{(n)}$. 
The pion is identified with the lowest pseudo-scalar mode $\omega^{(0)}$.
Eq. (\ref{mass_pscalar}) shows that the pion acquires mass due to the
non-trivial tachyon background.
Below we will estimate this mass for the case of a small quark mass.

\subsection{The pion mass}

We will begin by establishing that the pion is massless when the quark
is massless. In this case 
the large $u$ behavior of the tachyon and the 8-brane-anti-8-brane
separation is given by
\be
T(u)\approx C_T u^{-2} e^{-RL_\infty u} \; , \;
L(u)\approx L_\infty + C_L u^{-9/2} \,,
\ee
where $C_T$ is related to the chiral condensate.
For $M_0=0$  the solution to eq. (\ref{eigen_zeta}) is
\be
\zeta_0(u) = 0 \,.
\ee
The large $u$ asymptotic form of $\xi^\parallel_0(u)$ can then be read off
from the asymptotic behavior of eq. (\ref{eigen_zetaxi}):
\be
\label{long_0_mode}
\xi^{\parallel}_0(u) \approx a + bu^{-3/2} \,,
\ee
where $a$ and $b$ are constants which we will determine shortly.
It can easily be checked that this solution is UV normalizable under the
condition (\ref{norm_xipara}), and therefore that it corresponds to a massless
pion.

Now turn on a small quark mass $C^\prime_T$.
The condition for the validity of the corresponding tachyon solution
\be
T(u)\approx u^{-2} \left(C_T e^{-RL_\infty u} + C'_T e^{RL_\infty u}\right)\,,
\ee
is $C^\prime_T \lesssim e^{-RL_\infty u_\infty}$.
We can therefore treat
the quark mass as a perturbation of the massless solution we found above. 
To leading order in $C'_T$, and at large $u$, equation (\ref{eigen_zeta}) gives 
\be
\zeta_0(u) \approx -{3b M_0^2\over 8}
\left(C^\prime_Te^{RL_\infty u} + C_T e^{-RL_\infty u}\right)^{-2} \,,
\ee
where $M_0^2 = {\cal O}(C'_T)$.
We have kept the $C'_T$ term from the asymptotic solution for the tachyon
since it comes with a growing exponential.
We can now use (\ref{mass_pscalar}) to express the pion mass $M_0$ in terms
of the parameters $C_T$ and $C'_T$.
Since the integral in (\ref{mass_pscalar}) is 
dominated by large $u$ due to the presence of the tachyon potential,
we get
\be
M_0^{-2} \approx {9 b^2 {\cal N}\over 8 R^2}
\int^{u_\infty}_{u_1} {du\over\left(C^\prime_Te^{RL_\infty u} 
+ C_T e^{-RL_\infty u}\right)^2} \,,
\ee
where lower limit of the integral satisfies $1\ll u_1 \ll u_\infty$.
Under the condition that $C^\prime_T \lesssim e^{-RL_\infty u_\infty}$, and
assuming that $C_T\sim {\cal O}(1)$, this gives
\be
M_0^{-2} 
\approx  {9b^2{\cal N}\over 16 R^3 L_\infty C'_T C_T} \,.
\ee
Using (\ref{quark_mass}) and (\ref{C_T-qq}), and inverting we get
\be M_0^2 \approx
{8 R^4\over 9{\cal N}^2 b^2} {m_q \langle\bar{q}q\rangle} \,.
\ee

Let us now compute the constants $a$ and $b$ that appear
in the longitudinal zero mode (\ref{long_0_mode}) for zero quark mass.
In principle, these constants are fixed by the boundary conditions and the
normalization condition (\ref{norm_xipara}). However the latter requires
more knowledge than we have about the form of the solution at finite $u$.
Instead, what we will show is that these constants are related to the 
gauge theory parameter $f_\pi$, the so-called pion decay constant.
To see how, recall that in QCD the pion decay constant can be extracted
from the two point function of the axial vector current in the massless quark limit
\be
f_\pi^2 = \Pi_A(0) = \int d^4 x\, \langle J_A^\mu(x) J_{A \mu}(0) \rangle  \,.
\ee
Using the usual holographic dictionary, we evaluate this
by varying the action of the solution twice with respect 
to the boundary value of the dual field $A^\perp_\mu$.
Fourier-transforming in spacetime, the part of the action (\ref{A-_action}) that
depends on this field becomes
\be
S[A^\perp] = {\cal N} \int {d^4p \over (2\pi)^2} du\, uV(T)\sqrt{\cal D}\left[
	{\cal B}_1 |A^{\perp\prime}_\mu|^2
	+ (p^2 + {\cal B}_2) |A^\perp_\mu|^2
	- {\cal B}_4 A^{\perp\prime}_\mu A^{\perp\mu}
\right] \,,
\label{Pperp_action}
\ee
where the coefficients were defined in (\ref{Bcoefficients}). 
The equation of motion reads
\be
\partial_u \left[ uV(T)\sqrt{\cal D}\left({\cal B}_1 A^{\perp\prime}_\mu 
- {1 \over 2}{\cal B}_4 A^\perp_\mu \right) \right]
= uV(T)\sqrt{\cal D}\left((p^2+{\cal B}_2) A^\perp_\mu 
- {1 \over 2}{\cal B}_4 A^{\perp\prime}_\mu \right) \,.
\label{eom_calA}
\ee
Evaluating the action on the equation of motion then gives
\be
S[A^\perp] &=& {\cal N}\int {d^4p \over (2\pi)^2} du
	{d \over du}\left[ uV(T)\sqrt{\cal D}
	\left({\cal B}_1 A^{\perp\prime}_\mu A^{\perp\mu}
	- {1 \over 2}{\cal B}_4| A^\perp_\mu|^2 \right)\right] \nonumber \\
&=& {\cal N} \int {d^4p \over (2\pi)^2}\, uV(T)\sqrt{\cal D}
	\left( {\cal B}_1 A^{\perp\prime}_\mu A^{\perp\mu} 
	- {1 \over 2}{\cal B}_4| A^\perp_\mu|^2 \right) 
	\bigg|_{u=u_\infty} \,.
	\label{axial_action_pzero}
\ee
The IR boundary $u=u_0$ doesn't contribute since
the tachyon potential $V(T)$ goes to zero exponentially. 
Consider the zero mode
$A^\perp_\mu(p^\mu,u) = \xi^\perp_0(u) a_\mu^{-(0)}(p^\mu)$,
and impose the boundary condition 
$\xi_0^\perp(u_\infty) = 1$. 
The pion decay constant is evaluated by varying with respect to 
$a^{-(0)}_\mu$ and imposing $p^2=0$:
\be
f_\pi^2 =  {\cal N} uV(T)\sqrt{\cal D}
	\left( {\cal B}_1\xi_0^\perp(u) \xi_0^{\perp\prime}(u) 
	- {1 \over 2}{\cal B}_4\xi_0^\perp(u)^2 \right) \bigg|_{u=u_\infty} \,.
\label{decay_constant}
\ee
To relate this to the longitudinal zero mode we note that the equation
for $\xi^\perp_0(u)$ which follows from (\ref{eom_calA}) at $p^2=0$
is precisely the same as the equation for $\xi^\parallel_0(u)$ 
(\ref{eigen_zetaxi}) with $\zeta_0(u)=0$.
This implies that the two zero modes are proportional to each other.
The proportionality factor can be determined from the normalization 
condition for $\xi^\parallel_0(u)$ (\ref{norm_xipara}), which when combined 
with (\ref{eigen_zetaxi}) gives
\be 
{1 \over 2}={\cal N} u V(T) \sqrt{\cal D} 
\left({\cal B}_1 \xi^\parallel_0(u) \xi^{\parallel\prime}_0(u)
- {1\over 2}{\cal B}_4 \xi^\parallel_0(u)^2\right)\Bigg|_{u=u_\infty}\,. 
\label{normparaUV}
\ee
Comparing with (\ref{decay_constant}) we see that 
$\xi^\perp_0(u) = \sqrt{2}f_\pi\xi^\parallel_0(u)$.
The boundary condition on the longitudinal zero-mode  is therefore 
$\xi^\parallel_0(u_\infty)=1/(\sqrt{2}f_\pi)$,
and the condition (\ref{normparaUV}) then fixes the other constant:
\be
\xi^\parallel_0(u) \approx 
{1 \over \sqrt{2}f_\pi}-{\sqrt{2} f_\pi R^2 \over 3 {\cal N}}u^{-3/2} \,.
\label{xiUVapprox}
\ee
We finally get 
\be 
M_0^2 \approx {4 m_q \langle\bar{q}q\rangle \over f_\pi^2}\,,
\ee
which up to a factor of two reproduces the GOR relation.

\section{Conclusions}

The Sakai-Sugimoto model provides a holographic description of a gauge 
theory that is close to large $N_c$ QCD with massless quarks.
The most compelling feature of this model is that it exhibits spontaneous 
chiral symmetry breaking in a simple geometrical way.
We have extended this model by adding the flavor bi-fundamental scalar field
corresponding to the open strings between the 8-branes and anti-8-branes.
This field is dual to the operator $\bar{q}q$, and therefore describes
both the quark mass deformation, as well as the chiral-symmetry-breaking
quark condensate.

Our analysis was carried out in the non-compact AHJK model,
which is really dual to the NJL model.
However, since the UV behavior in the compact and non-compact models 
is identical, our results for the quark mass, quark condensate, and GOR relation
hold in the compact Sakai-Sugimoto model as well.
The IR behavior of the solution will be quantitatively different, though
qualitatively similar. 

In the non-compact model
there exists also a parallel 8-brane-anti-8-brane configuration in which the
chiral symmetry is unbroken.
In this configuration the bi-fundamental field is tachyonic near the origin,
so the configuration is locally unstable. The stable configuration is the connected, U-shaped
configuration in which the chiral symmetry is broken.
The same two configurations exist in the Sakai-Sugimoto (compact) model
at high temperature. It would be interesting to study the effect of
the bi-fundamental scalar in that case as well.
In particular if the tachyonic mode becomes massless at some temperature,
it might indicate a second-order (rather than first-order) phase transition
to a chiral-symmetric phase. 

Far from the origin the bi-fundamental scalar field is massive, but it still
has an important role in the holographic duality.
The 8-brane-anti-8-brane theory with this field exhibits a U-shaped
8-brane solution with a non-trivial profile for the bi-fundamental field.
At large distance (the UV of the gauge theory) the bi-fundamental field
has a non-normalizable exponentially growing component, 
and a normalizable exponentially decreasing component. 
We showed that the former is related to the quark mass, and that the latter
is related to the quark condensate.
We have also found a numerical solution in the normalizable case,
and compared it to the solution of the model without the bi-fundamental field.
We showed that including this field lowers the free energy of the solution. 

Lastly, we began an analysis of the fluctuations of the 8-brane
and anti-8-brane worldvolume fields, including the bi-fundamental
scalar, and the adjoint scalar and gauge fields. 
These correspond to the various mesons.
In particular the lowest mode of the $u$-component of the antisymmetric
combination of the gauge fields describes the pseudo-scalar pion.
We evaluated the mass of this mode in terms of the asymptotic behavior of
the bi-fundamental field, and showed that it satisfies the GOR relation.
In other words the pion mass is proportional to the product of the quark mass
and the quark condensate. The rest of the meson spectrum should
be analyzed numerically.


There is an interesting question regarding the bi-fundamental field.
In the action (\ref{Garousi_action}) the field $T$ corresponds to a straight
string along $x_4$. However, as was shown in \cite{Antonyan:2006vw},
the open string between the 8-brane and anti-8-brane curves into a U-shape.
It is not immediately clear to us how to incorporate this effect into the action.
It would be interesting to study to what extent this affects the solution with $T$.

Another interesting open question is the spectrum of fluctuations of the bi-fundamental
scalar field $T$.
In particular the spectrum of fluctuations around the parallel and U-shaped 
solutions with $T=0$ will determine their (in)stability.

\acknowledgments
We would like to thank Ofer Aharony, Roberto Casero, 
Nadav Drukker, Nissan Itzhaki, Elias Kiritsis,
Gilad Lifschytz, Akitsugu Miwa, Angel Paredes, and Shigeki Sugimoto
for fruitfull discussions.
We thank Ofer Aharony for his comments on the manuscript.
The work of S.S. and J.S. was supported in part by the Israel Science Fundation, by  a grant
(DIP H52) of German Israel Project Cooperation and by the European Network MRTN-CT-2004-512194.
O.B. is supported in part by the
Israel Science Foundation under grant no.~568/05.


\begin{thebibliography}{99}

\bibitem{Sakai:2004cn}
  T.~Sakai and S.~Sugimoto,
  Prog.\ Theor.\ Phys.\  {\bf 113}, 843 (2005)
  [arXiv:hep-th/0412141].
  
\bibitem{Witten:1998zw}
  E.~Witten,
  Adv.\ Theor.\ Math.\ Phys.\  {\bf 2}, 505 (1998)
  [arXiv:hep-th/9803131].
  
\bibitem{GellMann:1968rz}
  M.~Gell-Mann, R.~J.~Oakes and B.~Renner,
  Phys.\ Rev.\  {\bf 175}, 2195 (1968).

\bibitem{Casero:2005se}
  R.~Casero, A.~Paredes and J.~Sonnenschein,
  JHEP {\bf 0601}, 127 (2006)
  [arXiv:hep-th/0510110].
  
  
\bibitem{Aharony:2006da}
  O.~Aharony, J.~Sonnenschein and S.~Yankielowicz,
  arXiv:hep-th/0604161.
  
\bibitem{Peeters:2006iu}
  K.~Peeters, J.~Sonnenschein and M.~Zamaklar,
  Phys.\ Rev.\  D {\bf 74}, 106008 (2006)
  [arXiv:hep-th/0606195].



\bibitem{Peeters:2005fq}
  K.~Peeters, J.~Sonnenschein and M.~Zamaklar,
  JHEP {\bf 0602}, 009 (2006)
  [arXiv:hep-th/0511044].






  
\bibitem{Kruczenski:2003uq}
  M.~Kruczenski, D.~Mateos, R.~C.~Myers and D.~J.~Winters,
  JHEP {\bf 0405}, 041 (2004)
  [arXiv:hep-th/0311270].
  
  
\bibitem{Evans:2007jr}
  N.~Evans and E.~Threlfall,
  arXiv:0706.3285 [hep-th].
  
  
  
  
\bibitem{Bergman:2006xn}
  O.~Bergman and G.~Lifschytz,
  JHEP {\bf 0704}, 043 (2007)
  [arXiv:hep-th/0612289].
  
  
  
\bibitem{Antonyan:2006vw}
  E.~Antonyan, J.~A.~Harvey, S.~Jensen and D.~Kutasov,
  arXiv:hep-th/0604017.
  
  
\bibitem{Kutasov:2005rr}
  D.~Kutasov,
  arXiv:hep-th/0509170.
  
\bibitem{Giveon:2007fk}
  A.~Giveon and D.~Kutasov,
  Nucl.\ Phys.\  B {\bf 778}, 129 (2007)
  [arXiv:hep-th/0703135].
  
\bibitem{Garousi:2004rd}
  M.~R.~Garousi,
  JHEP {\bf 0501}, 029 (2005)
  [arXiv:hep-th/0411222].
  
\bibitem{Sen:1999md}
  A.~Sen,
  JHEP {\bf 9910}, 008 (1999)
  [arXiv:hep-th/9909062].
  
  
\bibitem{Hashimoto:2007fa}
  K.~Hashimoto, T.~Hirayama and A.~Miwa,
  JHEP {\bf 0706}, 020 (2007)
  [arXiv:hep-th/0703024].
  
  \bibitem{Witten:1979vv}
  E.~Witten,
  Nucl.\ Phys.\ B {\bf 156}, 269 (1979).
  
  %
\bibitem{Veneziano:1979ec}
  G.~Veneziano,
  Nucl.\ Phys.\ B {\bf 159}, 213 (1979).
  
    
\bibitem{Armoni:2003fb}
  A.~Armoni, M.~Shifman and G.~Veneziano,
  Phys.\ Rev.\ Lett.\  {\bf 91}, 191601 (2003)
  [arXiv:hep-th/0307097].

  


  
  
  
  
\bibitem{Erlich:2005qh}
  J.~Erlich, E.~Katz, D.~T.~Son and M.~A.~Stephanov,
  Phys.\ Rev.\ Lett.\  {\bf 95}, 261602 (2005)
  [arXiv:hep-ph/0501128].
  
\bibitem{Da Rold:2005zs}
  L.~Da Rold and A.~Pomarol,
  Nucl.\ Phys.\  B {\bf 721}, 79 (2005)
  [arXiv:hep-ph/0501218].


\bibitem{Casero:2007ae}
  R.~Casero, E.~Kiritsis and A.~Paredes,
  arXiv:hep-th/0702155.

  
\bibitem{Garousi:2000tr}
  M.~R.~Garousi,
  Nucl.\ Phys.\  B {\bf 584}, 284 (2000)
  [arXiv:hep-th/0003122].
  

\bibitem{Sen:2003tm}
  A.~Sen,
  Phys.\ Rev.\  D {\bf 68}, 066008 (2003)
  [arXiv:hep-th/0303057].
  
  

\bibitem{Kutasov:2000aq}
  D.~Kutasov, M.~Marino and G.~W.~Moore,
  arXiv:hep-th/0010108.

\bibitem{Buchel:2002tj}
  A.~Buchel, P.~Langfelder and J.~Walcher,
  Annals Phys.\  {\bf 302}, 78 (2002)
  [arXiv:hep-th/0207235];
   A.~Buchel and J.~Walcher,
  Fortsch.\ Phys.\  {\bf 51}, 885 (2003)
  [arXiv:hep-th/0212150].

\bibitem{Leblond:2003db}
  F.~Leblond and A.~W.~Peet,
  JHEP {\bf 0304}, 048 (2003)
  [arXiv:hep-th/0303035].

\bibitem{Lambert:2003zr}
  N.~Lambert, H.~Liu and J.~M.~Maldacena,
  JHEP {\bf 0703}, 014 (2007)
  [arXiv:hep-th/0303139].
  
 
  

  
  
\end{thebibliography}
\end{document}